\documentclass[
	aps,
	twocolumn,
	altaffilletter,
	nolongbibliography,
	numerical,
	flushbottom,
	secnumarabic,
	pra,
	superscriptaddress,
	floatfix,
	10pt
]{revtex4-2}

\usepackage{lipsum}
\usepackage{graphicx,subfigure}
\usepackage{braket}
\usepackage{amsmath, amssymb}
\usepackage{newtxtext, newtxmath}
\usepackage{booktabs}
\usepackage{siunitx}
\usepackage{bm}
\renewcommand{\vec}{\bm}

\usepackage[breaklinks, pdftex, hyperfootnotes=true, pdfpagelabels, bookmarks, pageanchor]{hyperref}
\hypersetup{%
	colorlinks=true, pdfstartpage=1, pdfstartview=FitH, pdfborder={0 0 0},%
	breaklinks=true, pdfpagemode=UseNone, pageanchor=true, pdfpagemode=UseOutlines,%
	plainpages=false, bookmarksnumbered, bookmarksopen=true, bookmarksopenlevel=1,%
	hypertexnames=true, pdfhighlight=/O,
	urlcolor=blue, linkcolor=blue, citecolor=blue
}

\newcommand{\eg}    {e.\,g.}

\newcommand{\comm}    [2]{\left[#1, #2\right]}

\newcommand{\tf}{\tau}
\newcommand{\gap}{\Delta}
\newcommand{\mingap}{\gap_\text{min}}

\DeclareMathOperator{\tts}{TTS}

\date{\today}

\newcommand{\addition}[1]{#1}
\newcommand{\replace}[2]{#2}
\newcommand{\delete}[1]{ }

\begin{document}

\title{Counterdiabatic Reverse Annealing}

\author{Gianluca Passarelli}
\email{gianluca.passarelli@spin.cnr.it}
\affiliation{CNR-SPIN, c/o Complesso di Monte S. Angelo, via Cinthia - 80126 - Napoli, Italy}

\author{Procolo Lucignano}
\affiliation{Dipartimento di Fisica ``E.\,Pancini'', Universit\`a di Napoli Federico II, Complesso di Monte S.~Angelo, via Cinthia - 80126 - Napoli, Italy}

\begin{abstract}
We present Counterdiabatic Reverse Annealing, a novel quantum annealing protocol that extends the range of application of reverse annealing to the previously inaccessible short-time domain. This is achieved by exploiting approximate counterdiabatic driving expanded in low-order nested commutators. In this work, we offer a comparative study of the performance of this new technique to that of unassisted reverse annealing in terms of metrics such as the ground-state fidelity and the time to solution. We provide a quantitative measure of the energetic cost of counterdiabatic reverse annealing and show that significant improvements are possible even using local counterdiabatic potentials, paving the way toward the experimental implementation in near-term quantum devices.
\end{abstract}

\maketitle

\section{Introduction}\label{sec:intro}

Adiabatic quantum computation (AQC) aims to find the ground state of an interacting classical Hamiltonian $H_P$ by exploiting quantum fluctuations and slow time evolutions~\cite{albash:review-aqc,Tanaka:book,Hauke:2019aa}. This powerful algorithm is used to tackle NP-hard optimization problems~\cite{Perdomo-Ortiz2012,Ramsey-expt,ronnow:speedup,Rieffel:2015aa,Azinovic:2016uq,Mott:2017aa,Li:comp-bio-2017,Mandra:2017ab,Jiang2018,Venturelli2019,Smelyanskiy:2018aa,Zlokapa:2019ab}, where the goal is to minimize a given cost function, since their solutions can always be mapped to ground states of spin-$1/2$ Hamiltonians~\cite{lucas:np-complete}. Rather than directly diagonalize $H_P$, which would require an exponentially growing computational effort as a function of the number of variables $N$, AQC seeks a continuous path of quantum states that connects the target ground state to a simpler initial state, such as the ground state $\ket{\psi_0^\text{TF}}$ of a noninteracting driver Hamiltonian $V_\text{TF}$. In standard AQC, the driver field has a twofold role: on the one hand, it allows to specify the starting state, on the other hand it provides the quantum fluctuations needed to escape the local optima of the cost function through tunneling. In fact, the driver Hamiltonian is nondiagonal in the computational basis and induces spin flips, providing the spins with the kinetic energy necessary to reach the global optimum~\cite{kadowaki:qa,farhi:quantum-computation}. To this end, a typical choice is
\begin{equation}\label{eq:transverse-field}
    V_\text{TF} = -\Gamma \sum_{i=1}^N \sigma_i^x,
\end{equation}
where $\sigma_i^x$ is the $x$ Pauli matrix acting on the $i$th spin of the system and $\Gamma$ is the strength of the transverse field. Its ground state $\ket{\psi_0^\text{TF}}$ is the equally-weighted quantum superposition of all the possible classical states of the computational basis and provides an unbiased starting state similar to the high-temperature starting state normally used in thermal annealing~\cite{simulated-annealing-review}.

The time-dependent Hamiltonian
\begin{equation}\label{eq:qa}
    H_\text{AQC}(s) = (1-s) V_\text{TF} + s H_P
\end{equation}
is used to drive the state $\ket{\psi_0^\text{TF}}$ to the target ground state $\ket{\psi_P} $ of $H_P$. In Eq.~\eqref{eq:qa}, $s = s(\theta)$ is a time-dependent external field, where $\theta = t/\tf \in [0,1]$ is a normalized time with respect to $\tf$, the total annealing time of evolution. The field $s$ is such that $s(0) = 0$ and $s(1) = 1$. The algorithm succeeds if $\tau$ is sufficiently long: the adiabatic theorem of quantum mechanics ensures that, during the evolution, the system will remain in the instantaneous ground state of $H_\text{AQC}(s)$, if there is a finite gap $\gap(s)$ separating the ground state from the rest of the spectrum, and provided that $\tf \gg O(\max_s \Delta^{-1}(s)) \equiv O(\mingap^{-1})$~\cite{Jansen:07,mozgunov2020quantum}.

Small spectral gaps are the bottleneck of AQC. Many optimization problems feature exponentially vanishing minimum gaps $\mingap \sim \exp(-\alpha N)$, which in turn imply an exponentially long annealing time $\tau \sim \exp(\alpha N)$ to satisfy the adiabatic criterion. This critical slowing down usually takes place when the system undergoes a first-order quantum phase transition (QPT); by contrast, a second-order QPT yields a minimum gap closing polynomially as a function of $N$, hence the annealing time for adiabaticity grows mildly (\eg, as a power law) as a function of the system size~\cite{Sachdev-book}. Smart strategies aimed to turn first-order QPTs into second order ones are therefore highly desirable in the race towards quantum advantage.

In this context, there have been many proposals in recent years. The idea is that additional terms in the Hamiltonian of Eq.~\eqref{eq:qa} may allow avoidance of first-order quantum phase transitions in the thermodynamic limit. Finite-size systems typically also benefit from the improved scaling of the minimum gap with $N$, which is important in the noisy-intermediate scale quantum era, where the quantum volume at our disposal is limited. Alternatively, the annealing path itself can be modified to improve performance. Among the possible strategies analyzed in the literature, we mention nonstoquastic auxiliary Hamiltonians~\cite{nishimori:non-stoq,seki:non-stoq,nishimori:non-stoq-2,seoane:transverse-interactions}, which are however still object of discussion~\cite{Crosson2020designing,imoto:catastrophic-nonstoq}, longitudinal or inhomogeneous transverse driving fields~\cite{PhysRevLett.123.120501,nishimori:inhomogeneous-1,nishimori:inhomogeneous-2,Adame:2020aa,albash2021diagonal}, optimal annealing paths~\cite{roland-cerf,susa:variational-lhz,takahashi:sta-qa,hegde:genetic,hegde:deep-learning-schedules} which may or may not include mid-anneal pauses~\cite{marshall,izquierdo2020ferromagnetically,passarelli:pausing,PhysRevApplied.14.014100}, and reverse annealing~\cite{perdomo:sombrero,chancellor:reverse,nishimori:reverse-pspin,yamashiro:ara,passarelli:reverse-ira,passarelli:reverse-ara,Venturelli2019,Ikeda2019}. In particular, reverse annealing is based on the idea that it is possible to initialize the algorithm in a state that is already similar to the target, often speeding up convergence.

A different approach that also exploits auxiliary terms in the Hamiltonian is shortcuts to adiabaticity across small spectral gaps~\cite{delcampo:assisted-adiabatic-passage,torrontegui:sta,guery-odelyn:sta,campbell:sta,hatomura:sta-classical-qa,delcampo:quantum-engines,delcampo:qsl-oqs}, using counterdiabatic driving (CD)~\cite{delcampo:sta-cd}. A counterdiabatic driving potential entirely suppresses nonadiabatic transitions arising from finite-$\tf$ sweeps and replicates adiabatic dynamics even if $\tf < O(\mingap^{-1}) $, effectively avoiding the problem of small gaps. Unfortunately, the price to pay is the inclusion in the Hamiltonian of a highly nonlocal operator whose norm, and consequently its energetic cost, diverges at the quantum critical point~\cite{demirplack:2008,berry:cd}. Dealing with this problem entails renouncing\delete{to}an exact CD operator in favor of an approximate one, which can be built with the aid of a variational principle~\cite{polkovnikov:pnas} and can be more easily implemented experimentally. Previous works have shown that approximate counterdiabatic driving are helpful for many-body quantum systems~\cite{passarelli:cd-pspin,hartmann:lhz,xie:cd-hubbard,ozguler:random-spins}.

In classical optimization, it is known that combining together different heuristic optimization methods can help improve the quality of the solution compared to each method alone~\cite{modern_optimization_book}. Motivated by this idea, in this paper we aim to study the combined effects of approximate CD driving and reverse annealing in order to exploit the benefits of both strategies in a single protocol, which we name counterdiabatic reverse annealing (CRA). Reverse annealing, in its adiabatic incarnation (ARA)~\cite{nishimori:reverse-pspin}, can turn first-order QPTs into second-order ones if certain conditions are met, as we are going to elucidate. Our aim is to study if and how the removal of the pathological exponential closing of the gap with the system size $N$ does also affect the performance and the energetic cost of approximate CD driving. We will focus on a closed-system setting where dynamics are unitary. In this framework, we can expect that using CRA will allow us to get the best of both worlds, extending ARA to the previously inaccessible short-time domain.

This paper is organized as follows. In Sec.~\ref{sec:pspin-ara}, we introduce the model we are going to study and its general properties. Additionally, we describe the main strengths and pitfalls of the adiabatic reverse annealing protocol. In Sec.~\ref{sec:cd}, we discuss exact and approximate CD driving and its energetic cost. In Sec.~\ref{sec:results}, we combine CD and ARA into CRA and show the results of our numerical simulations. Finally, we draw our conclusions in Sec.~\ref{sec:conclusions}.

\section{p-spin model and adiabatic reverse annealing}
\label{sec:pspin-ara}

As an example of a system featuring a first-order QPT, we focus on the ferromagnetic $p$-spin model~ \cite{bapst:p-spin,gross:p-spin,derrida:p-spin}. The problem Hamiltonian reads
\begin{equation}\label{eq:pspin}
    H_P = -NE_0{\left(\frac{1}{N} \sum_{i=1}^N \sigma_i^z\right)}^p,
\end{equation}
where $E_0$ is the  energy scale of the problem. In the following, we will express energies as multiples of $E_0$ (and omit the latter for ease of notation). In addition, we will fix $\hbar = 1$, thus the “natural” time scale is $1/E_0$. For odd $p$, this Hamiltonian has a unique ground state, the ferromagnetic state with all spins pointing upwards: $\ket{\psi_P} = \ket{0,0,\dots,0}$. The resulting AQC Hamiltonian commutes with the square of the total spin operator, $S^2 = S_x^2 + S_y^2 + S_z^2$. This symmetry ensures that the orthogonal eigenspaces of the total spin operator are never mixed during the dynamics. Since the initial state $\ket{\psi_0^\text{TF}}$ and the target state $\ket{\psi_P} $ belong to the symmetry sector with maximum total spin, the dynamics never leaves this subspace of dimension $d = N + 1$.

During the anneal, the system undergoes a first-order QPT for $p \ge 3$ at a critical value of the driving field $s$, separating the initial paramagnetic phase from the target ferromagnetic phase. Even though the equilibrium properties of this model can be easily inferred using mean-field theory in the thermodynamic limit~\cite{bapst:p-spin}, the presence of a first-order QPT makes reaching the target state following an adiabatic evolution cumbersome. Hence, the $p$-spin system is a highly regarded benchmark for AQC and quantum annealing (QA)~\cite{passarelli:dissipative-p-spin,passarelli:genetic,wauters:pspin,passarelli:pausing,hegde:genetic,passarelli:reverse-ira,passarelli:reverse-ara,passarelli:cd-pspin,acampora:memetic,passarelli:cd-open,seoane:transverse-interactions,nishimori:reverse-pspin,nishimori:inhomogeneous-2,yamashiro:ara}. 

Using ARA, it is possible to tame the first-order QPT arising for $p \ge 3$. Adiabatic reverse annealing is a variant of AQC where the strength of quantum fluctuations varies nonmonotonically in time. As opposed to standard forward AQC, in ARA the system is prepared in the ground state of a classical Hamiltonian, supposed close in Hamming distance to the target state. If the distance to the target state is smaller than a certain threshold, a continuous path exists in the parameter space of the ARA Hamiltonian that does not cross any first-order critical line for every $p \ge 3$~\cite{nishimori:reverse-pspin}, yielding an exponential speedup of ARA compared to AQC in a closed system setting. This result was shown not to be robust with respect to weak local and global dephasing in the energy eigenbasis~\cite{passarelli:reverse-ara}, yet the exponential advantage provided by ARA in the nondissipative regime makes this technique compelling and worth exploring. The ARA Hamiltonian is the following:
\begin{equation}\label{eq:ara}
    H_\text{ARA}(\lambda, s) = (1-s)\lambda V_\text{TF} + (1-s) (1-\lambda) H_0 + s H_P.
\end{equation}
The key point is the addition of a new time-dependent external field $\lambda = \lambda(\theta)$ and an auxiliary Hamiltonian $H_0$. In the typical workflow of ARA, the evolution starts from $(\lambda(0), s(0)) = (0, 0)$ and ends in $(\lambda(1), s(1)) = (1, 1)$ following a smooth path $\lambda = \Lambda(s)$ in the parameter space that satisfies $\Lambda(0) = 0$ and $\Lambda(1) = 1$. In what follows, we are going to choose $\Lambda(s) = s^q$ with $q > 0$.

The term $H_0 = H_\text{ARA}(0, 0)$ is used to enforce a classical initial condition (i.\,e., its ground state) where $N_\uparrow$ spins are oriented upwards and the remaining $N_\downarrow = N - N_\uparrow$ are oriented downwards. In particular, we define
\begin{equation}\label{eq:h0}
    H_0 = -\sum_{i = 1}^N \epsilon_i \sigma_i^z = -\sum_{i\in I_\uparrow} \sigma_i^z + \sum_{i\in I_\downarrow} \sigma_i^z,
\end{equation}
where $\epsilon_i = \pm 1$ in units of $E_0$. When $\epsilon_i = 1$ ($\epsilon_i = -1$), the corresponding spin in the ground state $\ket{\psi_0}$ of $H_0$ is up-aligned (down-aligned) to minimize the energy. In Eq.~\eqref{eq:h0}, we grouped the spins according to the value of $\epsilon_i$, defining the set of indices such that $\epsilon_i = 1$ ($\epsilon_i = -1$) as $I_\uparrow$ ($I_\downarrow$). The initial state in the single-qubit computational basis is therefore
\begin{equation}
    \ket{\psi_0} = \Ket{\frac{1-\epsilon_1}{2}, \frac{1-\epsilon_2}{2}, \dots, \frac{1-\epsilon_N}{2}}.
\end{equation}
At $\theta = 1$, $H_\text{ARA}(1, 1) = H_P$. Therefore, an adiabatic evolution where the system is initialized in $\ket{\psi_0}$ will eventually converge to the target ground state of the problem Hamiltonian $H_P$ for sufficiently long annealing times. This must be contrasted with iterated reverse annealing (IRA), another realization of reverse annealing where, instead of adding terms to annealing Hamiltonian of Eq.~\eqref{eq:qa}, the annealing path $s(\theta)$ itself is modified, so that it goes from $s(0) = 1$ backwards up to an inversion point $s(\theta_\text{inv}) = s_\text{inv} < 1$ and then back again to $s(1) = 1$~\cite{perdomo:sombrero}. 

The new $H_0$ term breaks the spin symmetry of the original $p$-spin model. However, it is possible to show that the total spin operators of the two spin sets $I_\uparrow$ and $I_\downarrow$ are conserved quantities~\cite{nishimori:reverse-pspin}. Thus, the dynamics take place in the tensor product space of the two subspaces with maximum spins relative to the two spin sets. The dimension of this space is $d = (N_\uparrow + 1) (N_\downarrow + 1)$.
The supposed advantage of ARA comes from the fact that part of the system is already in the correct state and the algorithm only needs to flip $N_\downarrow$ spins. Since QPTs are collective phenomena, if a sufficiently large portion of the $p$-spin system is already in the correct state, then intuitively the QPT will not take place or, at least, it will be less abrupt~\cite{nishimori:reverse-pspin}. 

It is important to note that the order of the spin indices in the two sets $I_\uparrow$ and $I_\downarrow$ in Eq.~\eqref{eq:h0} is irrelevant. As a consequence, only the fraction of spins that are up-aligned in the initial state matters, but not their indices. We define this fraction as 
\begin{equation}
    c = N_\uparrow / N.
\end{equation}
This parameter is related to the initial magnetization along the $z$-axis [$m_0^z = (2/N) \braket{\psi_0 \vert S_z \vert \psi_0} = 2 c - 1$] and to the Hamming distance to the target ferromagnetic state [$h = N(1 - c)$]. We can also express the Hilbert space dimension as $d = (Nc + 1)[N(1-c) + 1] $, where we assume that $Nc$ is an integer. The maximum dimension for a system size $N$ is when $c = 1/2$ [$d = {(N/2 + 1)}^2 $], corresponding to a starting state with an equal number of up and down spins. Since in this case the starting state is effectively a random guess of the true solution, this is an unfavorable scenario for ARA. In this work, we will consider higher values of $c$.

\begin{figure}[t]
    \centering
    \includegraphics[width=0.6\columnwidth]{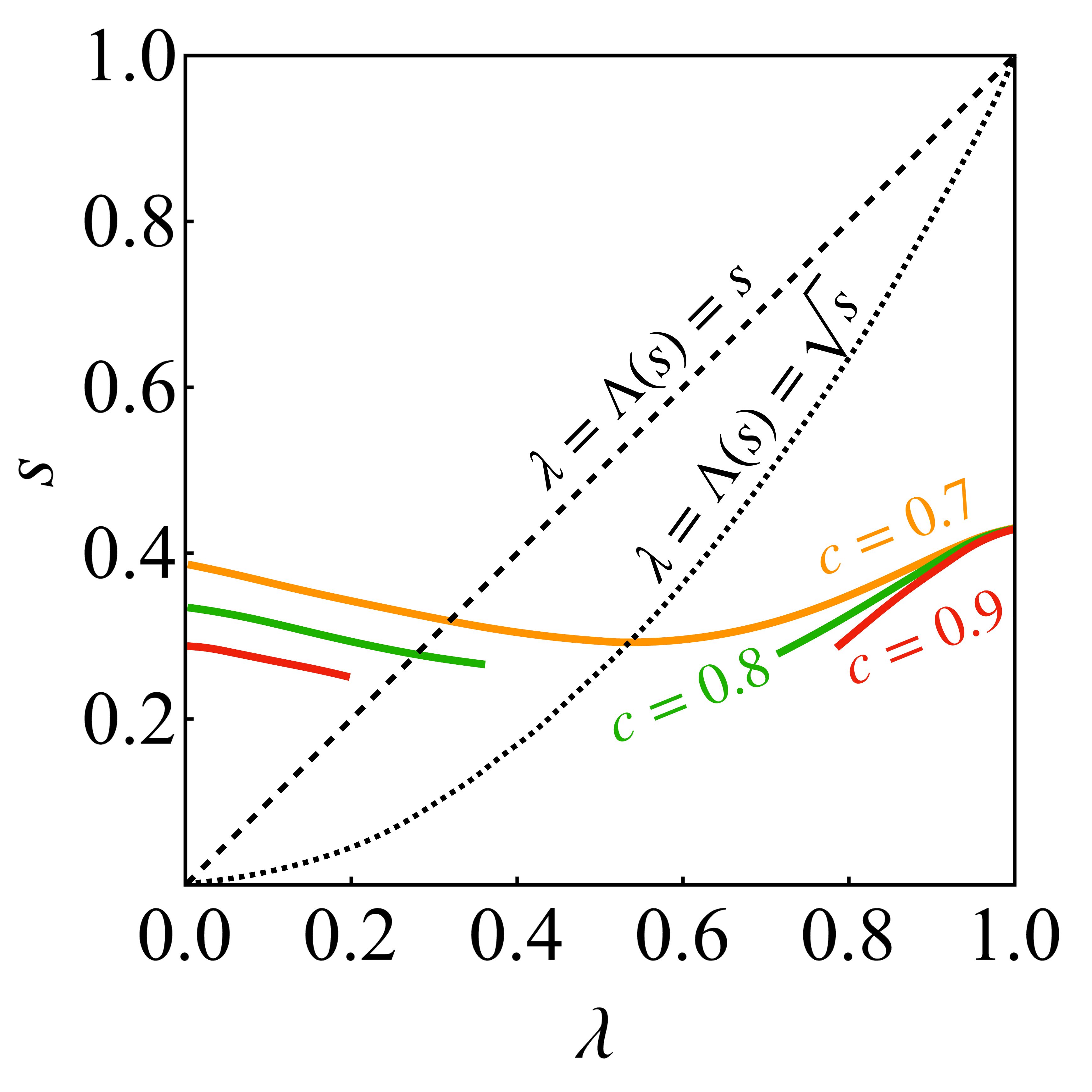}
    \caption{First-order critical lines in the $(\lambda, s)$ parameter space for several values of $c$. Parameters are $p = 3$ and $\Gamma = 1$. For $c > c^* = 0.74$, a gap in the middle opens up and allows reaching the final Hamiltonian without crossing a first-order QPT. Adapted from Ref.~\cite{yamashiro:ara}.}
    \label{fig:first-to-second-ara}
\end{figure}

Using a static approximation in the thermodynamic limit, the authors of Ref.~\cite{nishimori:reverse-pspin} showed that, when $c > c^*$, the first-order critical line in the $(\lambda,s)$ phase diagram splits: if a path $\lambda = \Lambda(s)$ is chosen that passes through the gap in the first-order critical line, the magnetization $m_z$ remains continuous and the quantum phase transition becomes second-order. A sketch of this effect is shown in Fig.~\ref{fig:first-to-second-ara} for $p = 3$. This effect also reflects on finite-size systems, where the scaling of the minimum gap with $N$ becomes polynomial when $c > c^*$ and, as a result, the scaling of the final-time error and the time to solution improve as well. The value of $c^*$ grows with increasing $p$; for $\Gamma = 1$ and $p = 3$, they find $c^* = 0.74$. The phase diagram for other values of $\Gamma$ and $p$ is qualitatively similar, but the gap opens up at different values of $c^*$. Some examples are discussed in Appendices~\ref{app:gamma-2} and~\ref{app:larger-p}.

The need for shortcuts to adiabaticity comes from the fact that ARA fails in the quench limit due to the breaking down of the adiabatic condition. On the other hand, we already know that approximate variational CD driving yields excellent performance on both the unitary and open system dynamics of the $p$-spin model~\cite{passarelli:cd-pspin,passarelli:cd-open}. Motivated by these results, our aim is to extend the regime of applicability of ARA to short annealing times exploiting approximate counterdiabatic driving.

\section{Counterdiabatic driving}
\label{sec:cd}

The idea of CD driving is to assist the adiabatic passage across small gaps exploiting an auxiliary term in the Hamiltonian~\cite{demirplack:2008,berry:cd,delcampo:assisted-adiabatic-passage}. The ARA Hamiltonian depends on time through the two-dimensional parameter vector $\vec{v} = (\lambda, s)$. Diabatic transitions between energy eigenstates of the parametric Hamiltonian $H(\vec{v}) = H_\text{ARA}(\lambda, s) $ are suppressed if the system is driven using the following Hamiltonian
\begin{equation}
    H_\text{CD}(\vec{v}) = H(\vec{v}) + \dot{\vec{v}} \cdot \vec{A}_{\vec{v}},
\end{equation}
where the overdot is the derivative with respect to time and $\vec{A}_{\vec{v}}$ is known as adiabatic gauge potential. When $\tf\to\infty$, the term proportional to $\dot{\vec{v}}$ goes to zero. The adiabatic gauge potential solves the following equation,
\begin{equation}\label{eq:gauge-equation}
    \comm{\vec{\nabla}_{\vec{v}} H(\vec{v}) + i \comm{\vec{A}_{\vec{v}}}{H(\vec{v})}}{H(\vec{v})} = \vec{0},
\end{equation}
whose solution involves the exact (unknown) energy eigenstates of $H(\vec{v})$. In many-body quantum systems, even in the rare cases where $\vec{A}_{\vec{v}}$ can be computed, the resulting operator is often nonlocal and its matrix elements diverge in the presence of QPTs, exponentially or polynomially in $N$ depending on the order of the transition. This singularity makes it impossible to adiabatically drive a quantum system across a point where the gap is zero or, in other words, the energetic cost of CD driving~\cite{campbell:cost-cd-1,campbell:cost-cd-2}, defined as the average of the Frobenius norm of the adiabatic gauge potential over the time domain,
\begin{equation}\label{eq:cost}
    \mathcal{C} = \frac{1}{\tf} \int_0^\tf \lVert \dot{\vec{v}} \cdot \vec{A}_{\vec{v}} \rVert \, dt,
\end{equation}
diverges if there are level crossings. Despite the divergence of $\mathcal{C}$ near closing gaps, CD driving can be useful to suppress transitions in finite-size systems where nonaccidental level crossings are always avoided, but the other issues remain, namely, nonlocality and requiring to know the spectral decomposition of $H$ to compute $\vec{A}_{\vec{v}}$. 

Despite these shortcomings, in recent years thanks to the assistance of a variational approach~\cite{polkovnikov:pnas} it has been possible to derive approximate CD protocols that are suitable for many-body systems~\cite{passarelli:cd-pspin,hartmann:lhz,xie:cd-hubbard,passarelli:cd-open}. In fact, solving Eq.~\eqref{eq:gauge-equation} is equivalent to minimizing the Frobenius norm of the operator
\begin{equation}\label{eq:operator-g}
    \vec{G}_{\vec{v}}(\vec{A}_{\vec{v}}^*) = \vec{\nabla}_{\vec{v}} H(\vec{v}) + i \comm{\vec{A}_{\vec{v}}^*}{H(\vec{v})}
\end{equation}
with respect to $\vec{A}_{\vec{v}}^*$. 
Reasonable \textit{ans\"atze} on $\vec{A}_{\vec{v}}$ can improve the ground state probability by several orders of magnitude compared with the case without corrections, without the need to access the full spectrum of $H(\vec{v})$. The compromise is that not all diabatic transitions are suppressed.

Some of these \textit{ans\"atze} are model-specific and/or work in specific sectors of the Hilbert space. For example, for the $S^2$-preserving AQC Hamiltonian of Eq.~\eqref{eq:qa}, a cyclic \textit{ansatz} yields excellent performance for the $p$-spin model with $p = 3$ with only $K = 3$ variational parameters~\cite{passarelli:cd-pspin}. On the other hand, an algebraic treatment of the adiabatic gauge potential~\cite{hatomura:algebraic-cd} leads to the following more general expansion in nested commutators (NCs),
\begin{equation}
    \vec{A}_{\vec{v}}^* = i \sum_{k = 1}^{K} \alpha_k(\vec{v}) [\underbrace{H(\vec{v}), [\dots,[H(\vec{v})}_\text{$2k-1$ times},\vec{\nabla}_{\vec{v}} H(\vec{v})]]]
\end{equation}
which is known to converge to the exact $\vec{A}_{\vec{v}}$ when $K \to \infty$~\cite{polkovnikov:nested-commutators}. Normally, we are interested in keeping only the low-order terms since, the larger is the order of the NCs, the less local is the resulting adiabatic gauge potential. This is why, in this manuscript, we consider up to $K = 3$.

The ARA Hamiltonian with the additional approximate CD driving terms gives rise to the counterdiabatic reverse annealing protocol. In the next section, we are going to study the performance of CRA for several choices of the Hamiltonian parameters. The order of approximation will be indicated as CRA$K$. 

\section{Results}
\label{sec:results}

Unless stated otherwise, we consider $q = 1$, that is, $\Lambda(s) = s$. While this is generally a significant restriction since it only allows to avoid the first-order phase transition if $c$ is very close to one~\cite{yamashiro:ara}, our aim is to assess the performance of CRA, with and without the first-order QPT, and it does not matter that we have to choose large values of $c$ to switch between the two cases. We can resort to a fine tuning of $\Lambda(s)$ in case the algorithm starts from a state with a smaller value of $c$. As for the annealing schedule, we consider the following polynomial function,
\begin{equation}\label{eq:schedule}
    s(\theta) = 6 \, \theta^5 - 15 \, \theta^4 + 10 \, \theta^3,
\end{equation}
whose first and second derivatives vanish at $\theta = 0$ and $\theta = 1$. At long times, this choice allows one to the reduce the exponent of the adiabatic error probability at the end of the computation thanks to boundary cancellation effects~\cite{morita:mathematical-foundation-qa,campos-venuti:boundary-cancellation}. In the context of this manuscript, it also means that the CD operator is zero at the boundaries since it is proportional to $\dot{\vec{v}}$. This is crucial since, at the beginning, the ARA Hamiltonian is employed to enforce the wanted classical starting condition, while, at the end of the annealing procedure, the Hamiltonian must be equal to the $p$-spin Hamiltonian whose ground state we want to reach.

We consider the $p = 3$ case. During our investigation, we also considered other values of $p$ and found results that are qualitatively and quantitatively similar to those shown in the main text (see also Appendix~\ref{app:larger-p}). We consider system sizes up to $N = 50$ and values of $c$ in the set $\set{0.7, 0.8, 0.9}$. Smaller values of $c$ are unfavorable since the starting state is too far from the target state and the QPT is always first-order. The case $c = 1$ is trivial since it means that the system is already in the target state before the start of the ARA algorithm. The annealing time is $\tf = 1$ in units of $1/E_0$ unless stated otherwise. Here, we consider $\Gamma = 1$ for the transverse field, for which the phase diagram of ARA is shown in Fig.~\ref{fig:first-to-second-ara}. In Appendix~\ref{app:gamma-2}, we also discuss the case of $\Gamma = 2$, which is qualitatively similar. We prepare the system in the ground state of $H_\text{ARA}(0, 0)$ and evolve it up to a time $ t = \tau \theta $ by propagating the state using $U(t) = \mathcal{T} \exp -i \, \tau \int_0^\theta H_\text{CD}(\lambda(\theta'), s(\theta')) \, d\theta'$, where $\mathcal{T}$ is the time ordering operator, and the counterdiabatic Hamiltonian is computed using the variational approach and the NC expansion for the ARA Hamiltonian $H_\text{ARA}(\lambda, s)$. We can then measure the probability that the system is in its ground state at the final time, $P_\text{GS} = {\lvert \langle \psi_P \vert U(t = \tau) \vert \psi_0 \rangle \rvert}^2$. 

\subsection{ARA vs CRA}
\begin{figure}[t]
\centering
\includegraphics[width=\columnwidth]{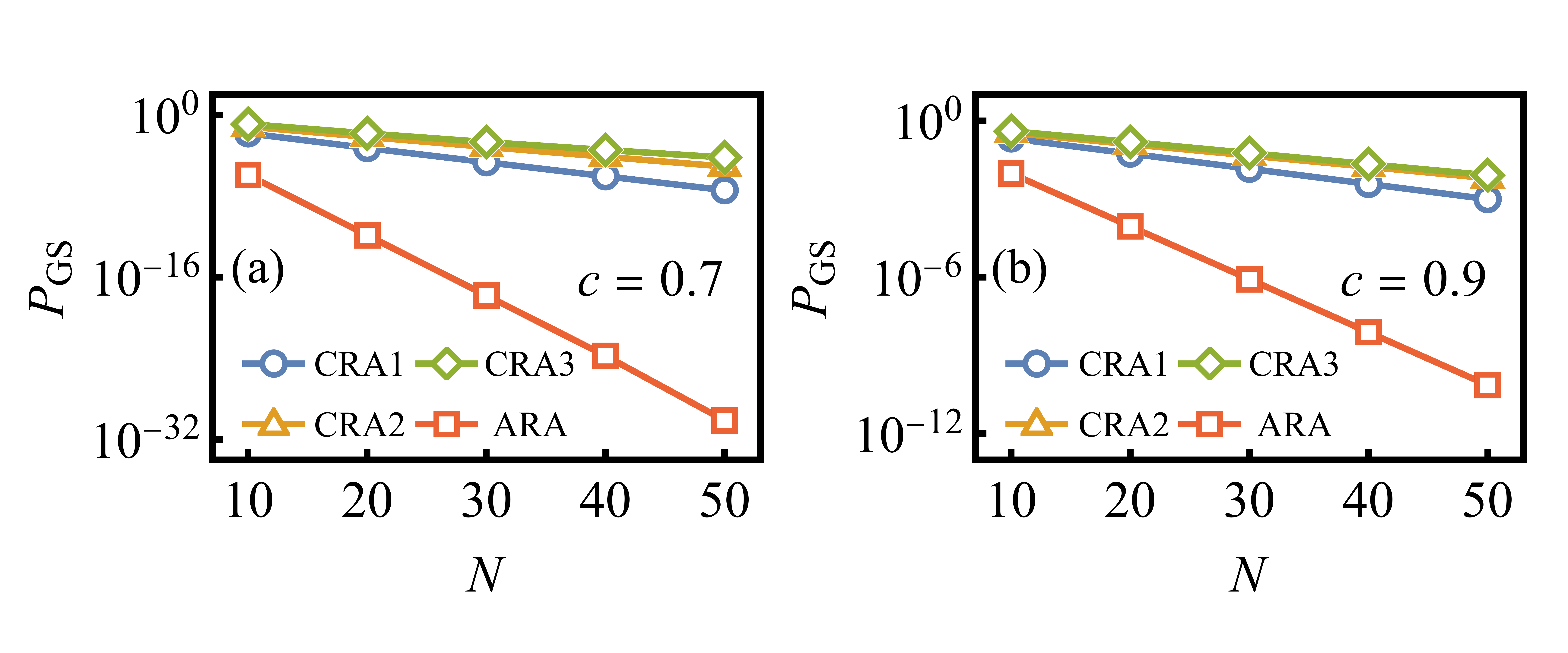}
\caption{Ground-state fidelity versus $N$, $p = 3$, for $\Lambda(s) = s$, $\Gamma = 1$, $\tau = 1$. Panel (a): $c = 0.7$. Panel (b): $c = 0.9$.}
\label{fig:c-0.7-0.9-p-3}
\end{figure}

In this subsection we compare the performance of ARA with that of CRA. In Fig.~\ref{fig:c-0.7-0.9-p-3}(a), we plot $P_\text{GS}$ as a function of the system size $N$ for $c = 0.7$. We observe that the ground-state fidelity when no CD corrections are applied decays exponentially. For $N=30$ spins, the fidelity is already practically zero, $P_\text{GS} \sim 10^{-18}$. By introducing approximate counterdiabatic corrections, the performance improves dramatically, by several orders of magnitude already for $K = 1$, where the scaling remains exponential but with a smaller exponent, see Tab.~\ref{tab:exponents}. By studying also $K = 2$ and $K = 3$, we see that the performance of CRA mostly saturates at $K = 2$ for small enough systems, which is worthwhile since it means that the locality of the approximate CD operator can be kept small, while still achieving satisfactory performance. In fact, for $N = 30$, we see that $P_\text{GS}^\text{CRA1} \sim 10^{-5}$ and $P_\text{GS}^\text{CRA2, CRA3} \sim 10^{-3}$. The exponents of the scaling functions are reported in Tab.~\ref{tab:exponents}.

\begin{table}[tb]
    \centering
    \begin{tabular}{l@{\hskip 2ex}c@{\hskip 2ex}c}
    \toprule
          & $c = 0.7$ & $c = 0.9$\\
    \midrule
    ARA & 2.00    & 0.68 \\
    CRA1   & 0.50    & 0.20 \\
    CRA2   & 0.34    & 0.15 \\
    CRA3   & 0.29    & 0.15 \\
    \bottomrule
    \end{tabular}
    \caption{Exponent $\gamma$ in the scaling law $P_\text{GS} \sim 2^{-\gamma N}$ for $c = 0.7$, $0.9$, with and without counterdiabatic corrections ($\Gamma = 1$).}
    \label{tab:exponents}
\end{table}

This saturation effect can also be seen in the case of $c = 0.9$, shown in Fig.~\ref{fig:c-0.7-0.9-p-3}(b), where instead the annealing path does not cross the first-order critical line. There are many similarities between Fig.~\ref{fig:c-0.7-0.9-p-3}(b) and Fig.~\ref{fig:c-0.7-0.9-p-3}(a): firstly, here as well the fidelity decreases exponentially as a function of $N$, although with a milder exponent than for $c = 0.7$; secondly, already a $K = 1$ CD operator improves the performance of ARA by several orders of magnitude, and, thirdly, the improvement yielded by CRA mostly saturates at $K = 2$, meaning that higher-order terms in the NC expansion only provide very small corrections. This is evident from Tab.~\ref{tab:exponents} since the scaling exponents for CRA2 and CRA3 are approximately the same. The main difference between the two panels of Fig.~\ref{fig:c-0.7-0.9-p-3} is the scale of the vertical axis: owing to the fact that the starting state is closer to the target for $c = 0.9$, the fidelity is significantly higher in this case (for instance, $P_\text{GS} \sim 10^{-6}$ for $N = 30$ already without CD corrections). By contrast, the spectral gap plays no role in this short-time regime. 

To see why, we can resort to perturbation theory for ARA without CD driving. Since the starting state has $N(1-c)$ down-aligned spins, the probability amplitude of flipping all the spins to the target state is $\sim \tilde{\Gamma}^{N(1-c)}$, where $\tilde{\Gamma}$ is an effective time-independent tunneling amplitude that ``summarizes'' the action of the time-dependent transverse field, given the fact that the dynamics are very fast and cannot see the fine details of the applied fields. It is reasonable to assume that $\tilde{\Gamma}$ is the mean value of $\Gamma \lambda (1-s)$ over the time domain, i.\,e., 
\begin{equation}\label{eq:integral}
    \tilde{\Gamma} = \tf \Gamma \int_0^1 \lambda(\theta) [1-s(\theta)] d\theta.
\end{equation}
For the annealing path with $\lambda(\theta) = s(\theta)$ [Eq.~\eqref{eq:schedule}], this integral yields $\tilde{\Gamma} = \tf \Gamma 25 / 231$. Thus, the probability of finding the system in its ground state is 
\begin{equation}\label{eq:pgs-perturbative}
    P_\text{GS}\sim 2^{2 \log_2(\Tilde{\Gamma}) (1-c) N},
\end{equation}
which is in very good agreement with our numerical results for all values of $c$ and $q$ in ARA. The addition of CD driving \addition{qualitatively} fits in this description as it increases the rate of spin flips and thus decreases the scaling exponent of the ground-state fidelity with $N$ in CRA\addition{, even though a full analytical calculation to compute $P_\mathrm{GS}$ would require the evaluation of the coefficient of the $l$th order in nested commutators, which is normally only obtained with numerical variational minimization}.

\begin{figure}[t]
    \centering
    \includegraphics[width=\columnwidth]{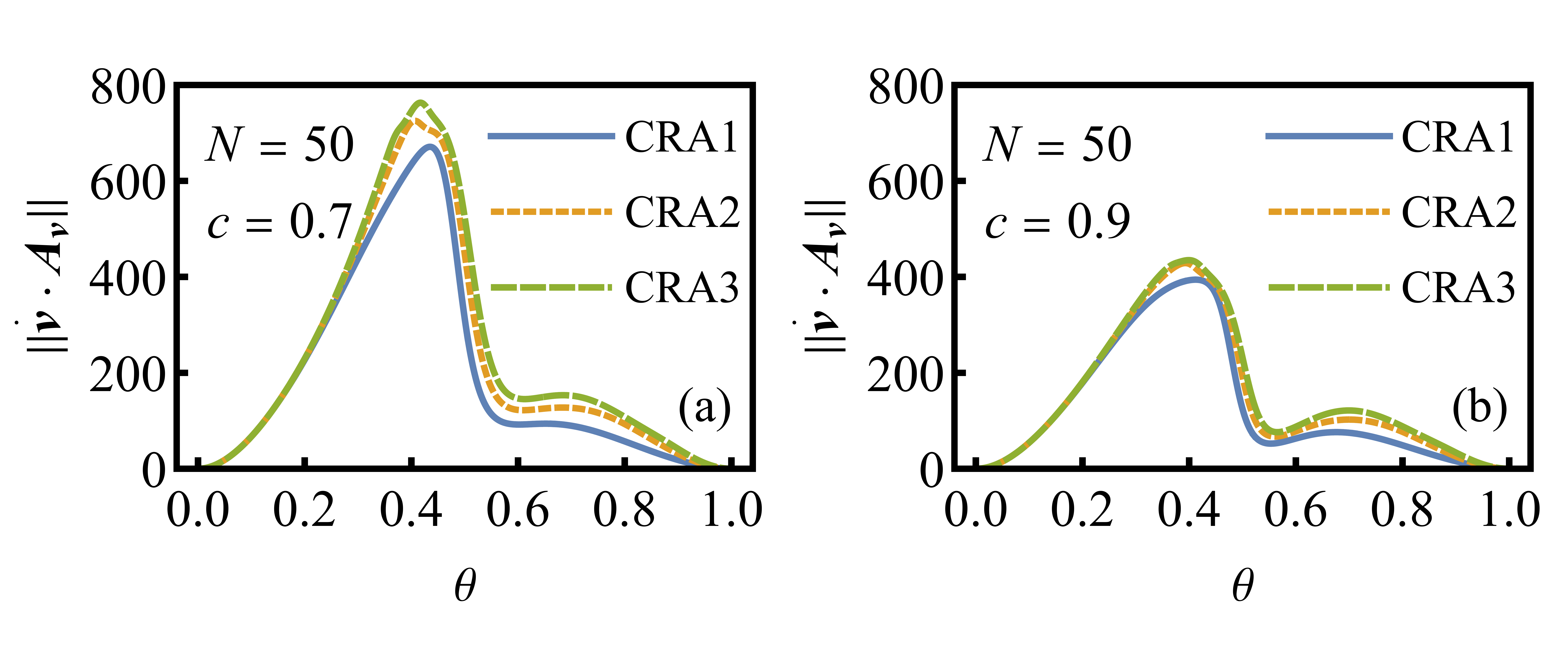}
    \caption{Norm of the counterdiabatic operator, in units of $E_0$, for $N = 50$. Here, $p = 3$ and $\Gamma =1$. Panel (a): $c = 0.7$; Panel (b): $c = 0.9$.}
    \label{fig:norm}
\end{figure}

The difference in performance between the two values of $c$ also reflects on the norm of the approximate CD operator, which is noticeably smaller for $c = 0.9$ compared to $c = 0.7$ for all orders of expansion in nested commutators. By studying the trace norm $\lVert \dot{\vec{v}} \cdot \vec{A}_{\vec{v}}\rVert$, we also notice that most of the contribution to the norm comes from the first order in the NC expansion (CRA1), as shown in Fig.~\ref{fig:norm} for $N = 50$. The tallest peak is found at the time of the minimum gap between the ground and first excited states. The tallest peak for $c = 0.7$ [panel (a)] is almost twice as high as the tallest peak for $c = 0.9$ [panel (b)]. The smaller peak corresponds to the CD operator suppressing transitions between excited states.

\begin{figure}[t]
    \centering
    \includegraphics[width=0.9\columnwidth]{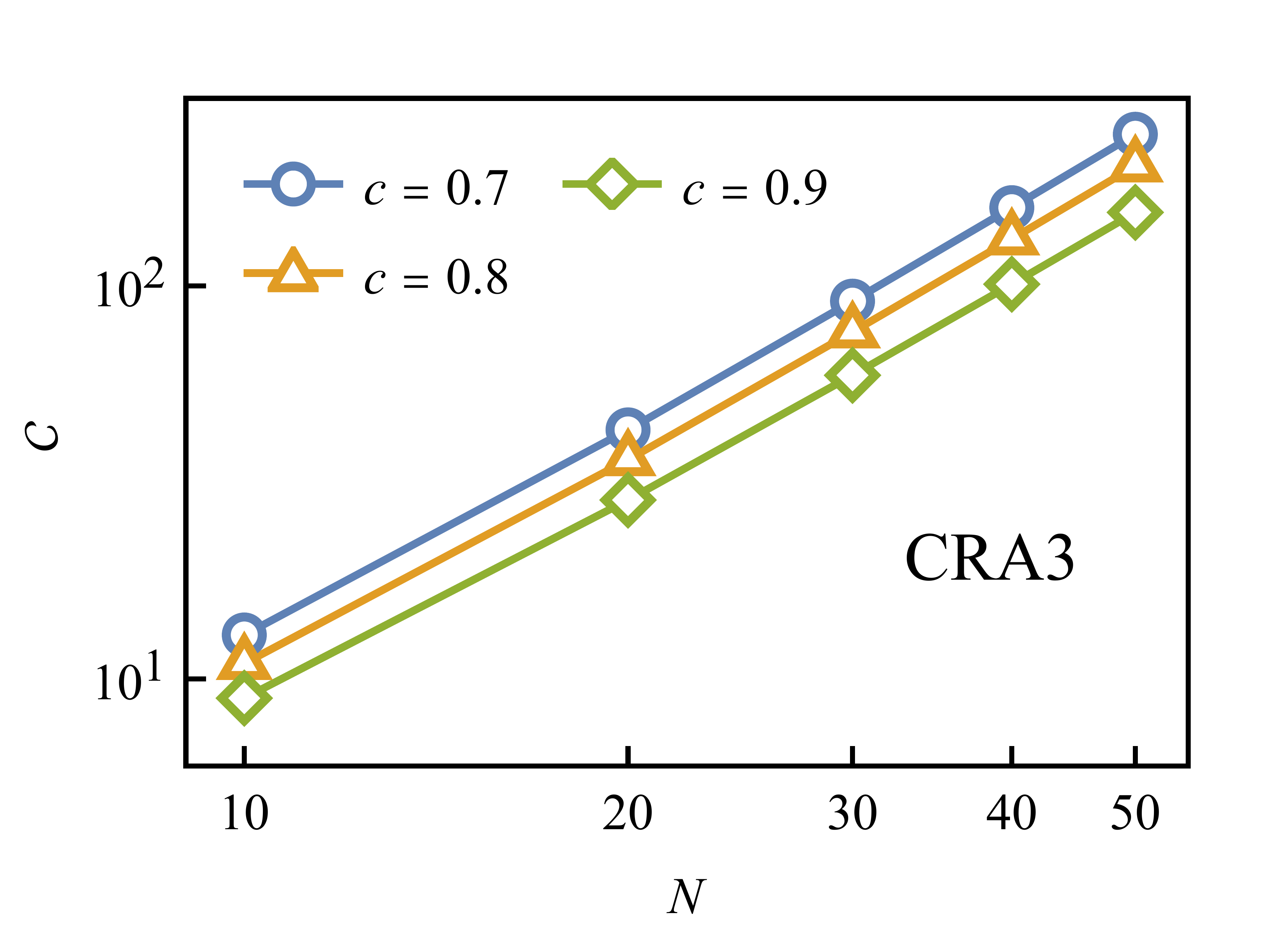}
    \caption{Energy cost of CD driving [Eq.~\eqref{eq:cost}], in units of $E_0$, as a function of $N$, for different values of $c$. Here, $p = 3$ and $\Gamma = 1$.}
    \label{fig:cost}
\end{figure}

The area below the plots in Fig.~\ref{fig:norm} represents the cost of counterdiabatic driving, Eq.~\eqref{eq:cost}. Intuitively, one would expect that the cost of a CD protocol is higher when the minimum gap is smaller, since in this case more effort is needed to drive the system adiabatically. \replace{By contrast, the scaling of the cost $\mathcal{C}$ with the system size $N$ in the short-time regime is not affected by the scaling of the gap with $N$ for the reasons discussed above.}{By contrast, due to finite-size effects and to the average over the short time domain, the scaling with $N$ of the cost of CD driving seems not to be affected by the order of the phase transition, despite earlier findings~\cite{hatomura:algebraic-cd}.} In fact, we report in Fig.~\ref{fig:cost} the scaling of the cost $\mathcal{C}$ in the case of CRA3 as a function of $N$ for three values of $c$, in a bilogarithmic scale. We can see that in all cases the cost scales as a power law $ \mathcal{C} \sim N^\alpha $, with an exponent that is close to $\alpha = 1.85$. The value of $c$ only determines a vertical offset.

\begin{figure}[t]
    \centering
    \includegraphics[width=\columnwidth]{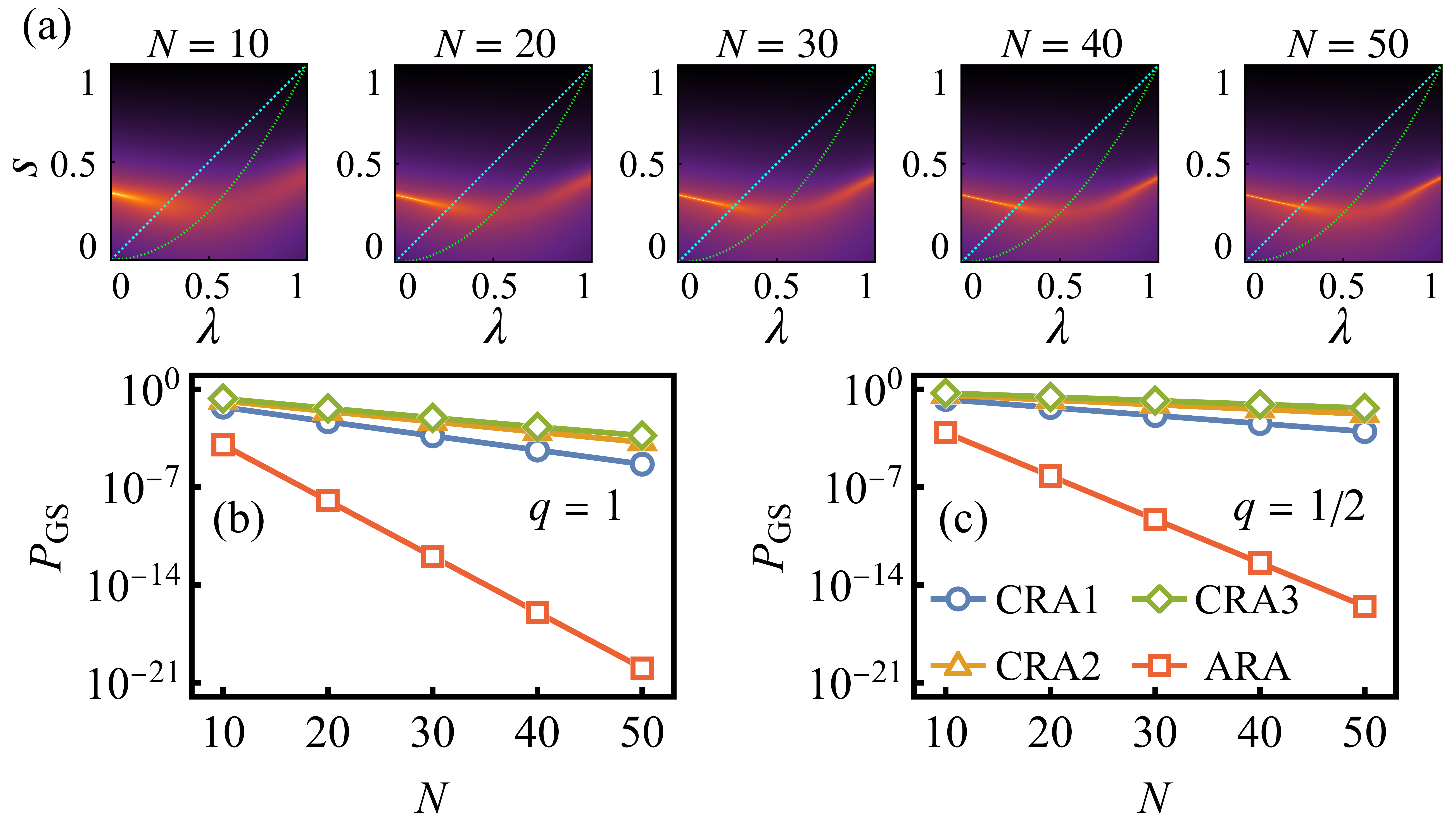}
    \caption{Panel (a): Cartoon of the inverse gap between the instantaneous ground and first excited state, $\Delta^{-1}$, as a function of $\lambda$ and $s$, for $c = 0.8$, $p = 3$, and $\Gamma = 1$. Brighter colors correspond to smaller gaps. Values have been rescaled so that they fall in the interval $[0,1]$ to improve visual clarity. For large enough systems ($N\gtrsim 30$), these plots closely resemble the ARA phase diagram of Fig.~\ref{fig:first-to-second-ara}, and the path $\Lambda(s) = \sqrt{s}$ allows avoiding the region with the smallest gaps. For the smaller $N = 10$ system, the two annealing paths encounter similarly-sized gaps and the performance of ARA in the two cases is the same. Panels (b-c): Ground-state fidelity versus $N$, $c = 0.8$, $p = 3$, $\Gamma=1$. Panel (b): $q = 1$; Panel (c): $q = 1/2$. The legend is the same for the two panels.}
    \label{fig:c-0.8}
\end{figure}

Next, we focus our attention on $c = 0.8$. As evident from Fig.~\ref{fig:first-to-second-ara}, the annealing path $\Lambda(s) = s$, corresponding to $q = 1$, crosses the first-order critical line, while the annealing path $\Lambda(s) = \sqrt{s} $, corresponding to $q = 1/2$, does not. We should mention that, in the latter case, $\vec{A}_{\vec{v}}$ has diverging matrix elements around $\theta = 0$ ($\sim \theta^{-3/2}$), however the counterdiabatic correction $\dot{\vec{v}} \cdot \vec{A}_{\vec{v}}$ is well-behaved thanks to $\dot{s}$ being zero ($\sim \theta^{2}$) at $\theta = 0$ [Eq.~\eqref{eq:schedule}]. The phase diagram of Fig.~\ref{fig:first-to-second-ara} is valid in the thermodynamic limit and does not necessarily reflect the physics of finite-$N$ systems, where finite-size effects might suppress any benefits of one path over the other. Therefore, in Fig.~\ref{fig:c-0.8}(a), we plot a cartoon of $1/\Delta$, rescaled so that the values of each plot lies in the interval $[0,1]$, as a function of $\lambda$ and $s$, for $c = 0.8$, for several values of $N$. Brighter colors correspond to smaller gaps. On top of each diagram, we also plot the two annealing paths, $\Lambda(s) = s$ (dashed cyan) and $\Lambda(s) = \sqrt{s}$ (dotted green). We see indeed that the two paths cross similarly valued gaps if $N\lesssim 20$, while for larger systems ($N \gtrsim 30$) the scenario is similar to the thermodynamic limit, where the path with $q = 1/2$ avoids the region with smaller gaps. Thus, on the one hand, for finite but large systems, we expect the choice of the annealing path to be important at sufficiently long times. On the other hand, our short-time analysis suggests that the performance of ARA and CRA when the annealing time is short should only depend on the integral of Eq.~\eqref{eq:integral} and not on the details of the spectrum. 

\begin{figure*}[tb]
    \centering
    \includegraphics[width=0.9\textwidth]{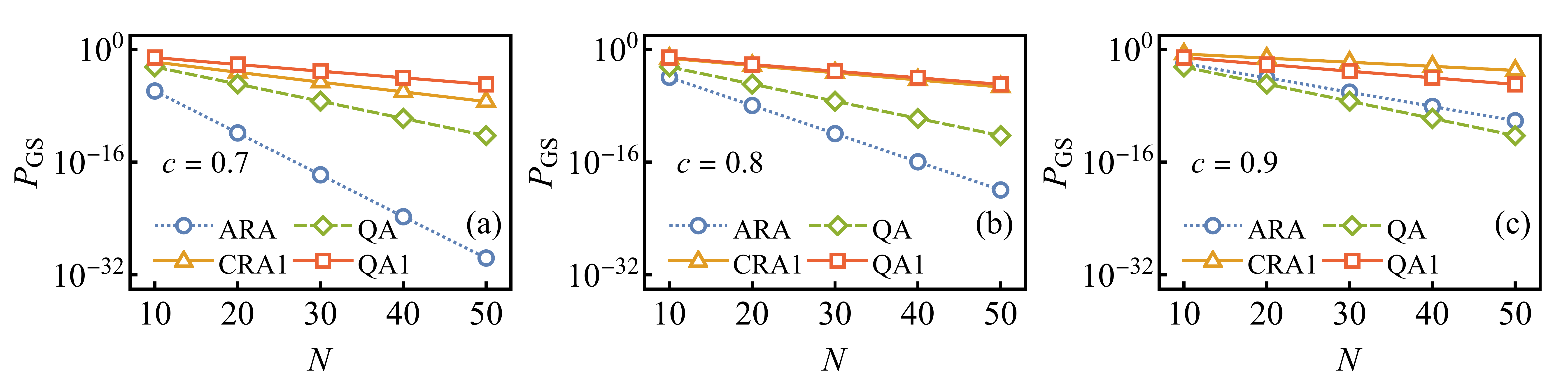}
    \caption{Ground-state fidelity of ARA and QA as a function of $N$, compared with the corresponding approximate counterdiabatic protocols (first-order in nested commutator expansion). The different panels refer to the initial state chosen in ARA and CRA. Panel (a): $c = 0.7$. Panel (b): $c = 0.8$. Panel (c): $c = 0.9$.}
    \label{fig:ara-vs-qa}
\end{figure*}

We numerically investigate this matter and show our results in Fig.~\ref{fig:c-0.8}(b-c). We can see that there is an advantage in using the annealing path $\Lambda(s) = \sqrt{s}$ as the decay of $P_\text{GS}$ becomes less steep, compared with the $\Lambda(s) = s$ path, both with and without counterdiabatic corrections. The exponents of the exponential scaling are shown in Tab.~\ref{tab:exponents-0.8}. The ARA exponents are consistent with our predictions based on Eq.~\eqref{eq:pgs-perturbative}. We also observe for the path $\Lambda(s) = \sqrt{s}$ the same features described before, i.\,e., near-convergence at the second order in nested commutators and a substantial improvement with just $K = 1$. We verified that the better performance of CD driving for $q = 1/2$ comes at the price of a larger norm of the resulting adiabatic gauge potential for any size $N$ (not shown).

\subsection{Comparison with standard QA}
In this subsection we show a comparison between adiabatic reverse annealing and standard quantum annealing with and without the CD correction. In both protocols, we consider a first-order expansion of the CD operator in nested commutators (CRA1 versus QA1). We consider $E_0 \tf = 1$ and compare the two protocols for all the values of $N$ and $c$ considered in this manuscript. We report our results in Fig.~\ref{fig:ara-vs-qa}, where we plot the ground state probability as a function of the number of spins. For the unassisted annealing protocols, we see that QA outperforms ARA unless the fraction $c$ is very close to one. For $c = 0.7$, we see that the ground-state fidelity of ARA falls short by several orders of magnitude compared to the one of standard QA. This fact expresses the main shortcoming of the ARA protocol, that is, it is essential to start from a very accurate initial state to for the algorithm to have good performance.

On the other hand, when we move to the assisted protocols, we see that the performance gap between reverse and forward annealing when $c = 0.7$ gets\delete{smaller}much smaller, showing that CD driving has a greater impact on ARA than on QA. In addition, for large values of $c$, CRA systematically performs comparably well or better than standard QA assisted by CD driving. Therefore, the inclusion of CD driving in the ARA protocol has a twofold advantage: on the one hand, it allows to extend the range of application of ARA to short times, and on the other hand it can compensate sub-optimal choices of the initial state of reverse annealing.

\begin{table}[tb]
    \centering
    \begin{tabular}{l@{\hskip 2ex}c@{\hskip 2ex}c}
    \toprule
          & $q = 1$ & $q = 1/2$\\
    \midrule
    ARA & 1.33    & 1.03 \\
    CRA1   & 0.35    & 0.20 \\
    CRA2   & 0.26    & 0.12 \\
    CRA3   & 0.22    & 0.09 \\
    \bottomrule
    \end{tabular}
    \caption{Exponent $\gamma$ in the scaling law $P_\text{GS} \sim 2^{-\gamma N}$ for $c = 0.8$, for $q = 1$ and $q = 1/2$ ($\Gamma=1$).}
    \label{tab:exponents-0.8}
\end{table}

%

\subsection{Time to solution}

Finally, we study the performance of CRA as a function of the annealing time for some values of $N$. One commonly used figure of merit is the so-called time to solution (TTS), which measures the effective time, in terms of number of repetitions of the same experiment, that one needs in order to get the exact solution at least once with a probability higher than a threshold $p_d$ (in the following, $p_d = 0.99$), given the $P_\text{GS}$ of each run of duration $\tf$~\cite{q108}. In formulas,
\begin{equation}\label{eq:tts}
    \tts(\tf) = \tf \frac{\log(1-p_d)}{\log(1-P_\text{GS})}.
\end{equation}
Typically, in the absence of CD corrections, the TTS is very large at short times, has a plateau when the dynamics is governed by Landau-Zener processes and $1-P_\text{GS} \sim \exp(-a\tf)$, and grows as $\tf / \log\tf$ at long times, where adiabatic perturbation theory predicts that the probabilities of the system being found in excited states is proportional to (powers of) $\tf^{-2}$~\cite{passarelli:reverse-ara}. This latter region is not visible in the window of annealing times that we chose to analyze in this section, hence the plots stop at the plateau. The region that is mostly affected by CD driving is the one at short times, where the term $\dot{\vec{v}} \cdot \vec{A}_{\vec{v}}$ is more relevant.

\begin{figure}[t]
    \centering
    \includegraphics[width=\columnwidth]{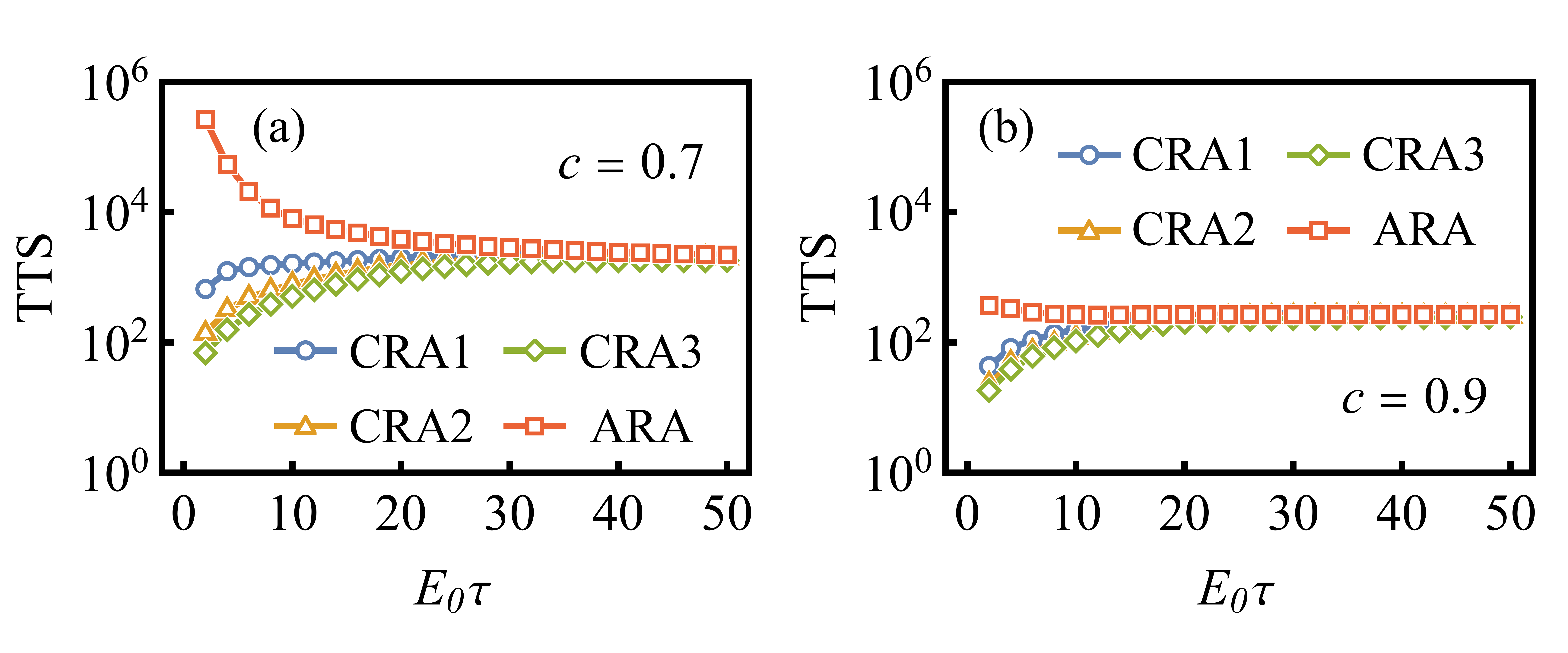}
    \caption{Time to solution\addition{, in units of $1/E_0$,} versus annealing time, for $N = 10$, $\Gamma = 1$, $q = 1$. Panel (a): $c = 0.7$; Panel (b): $c = 0.9$.}
    \label{fig:tts-10}
\end{figure}

\begin{figure}[t]
    \centering
    \includegraphics[width=\columnwidth]{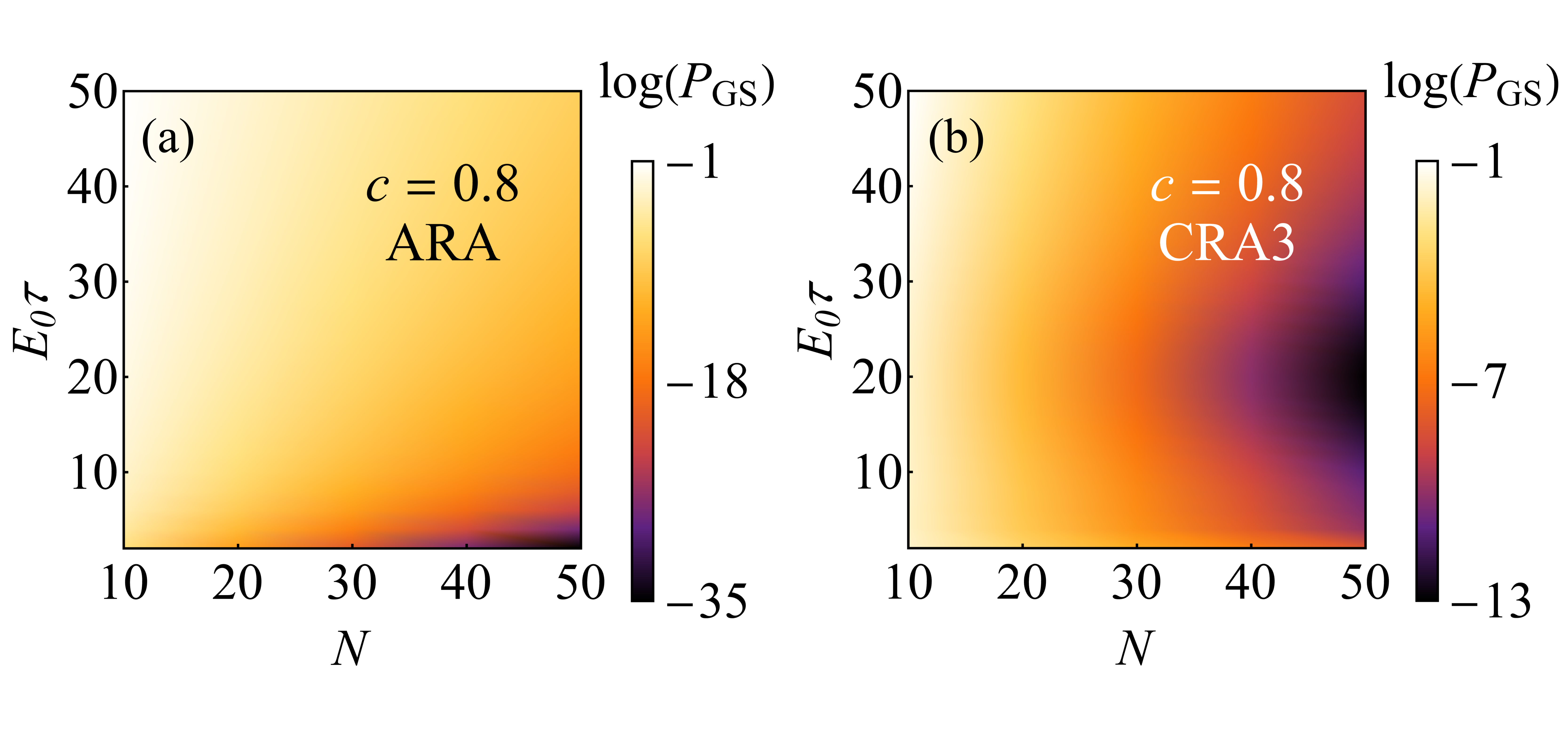}
    \caption{\addition{Natural logarithm of the} ground state probability versus system size $N$ and annealing time $\tf$, for $c = 0.8 $, $\Gamma = 1$, $q = 1$. Panel (a): ARA; Panel (b): CRA3. \addition{Note the different scale of the colorbars in the two panels.}}
    \label{fig:logpgs}
\end{figure}

In Fig.~\ref{fig:tts-10}, we show the TTS as a function of $\tf$ for $N = 10$ and $c = 0.7$ [panel (a)] and $c = 0.9$ [panel (b)] up to $\tf = 50$ in units of $1/E_0$. The annealing path is $\Lambda(s) = s$. We see that the CRA protocol is most effective at short times, where the improvement of the time to solution with respect to ARA can be quite large (almost by four orders of magnitude in the case of $c = 0.7$ when CRA3 is considered). Instead, at longer times the curves fall onto the one without CD corrections, as expected, since the effect of $\dot{\vec{v}} \cdot \vec{A}_{\vec{v}}$ becomes negligible when $\tf$ increases. Overall, we see that the TTS for $c = 0.9$ is shorter compared to $c = 0.7$, but we also notice that counterdiabatic driving yields a smaller relative improvement at short times in comparison to $c = 0.7$. We see qualitatively the same results also for other values of $N$. A summary of our analysis is shown in Fig.~\ref{fig:logpgs}, where we plot $P_\text{GS}$, in a logarithmic scale, as a function of both $N$ and $\tf$ in the case of $c = 0.8$. To make the plot smoother and easier to read, we linearly interpolated our results for the different $N$'s. In panel (a), we show the ground state probability when the CD correction is not applied: here, we see that $P_\text{GS}$ is very small at short times, according to the adiabatic theorem. When we apply CD driving, we see that the ground probability in this regime increases dramatically. We note that, at intermediate times, around $E_0 \tf \sim 25$, the performance of CD driving has a minimum as a function of $\tf$ for all system sizes. However, the $P_\text{GS}$ with CD driving is always higher than or equal to the ARA ground state probability.

\section{Conclusions}
\label{sec:conclusions}

In summary, we have introduced counterdiabatic reverse annealing, a heuristic protocol for quantum optimization combining adiabatic reverse annealing and variational counterdiabatic driving. We applied this technique to the ferromagnetic $p$-spin model, which is known to benefit from ARA in the unitary limit~\cite{nishimori:reverse-pspin,yamashiro:ara}. The main results of this paper are as follows. \textit{(i)}~We showed that CRA significantly improves ARA at short times for all values of the fraction of spin-ups that we considered. \textit{(ii)}~We showed that this improvement can be achieved almost entirely by using a first-order expansion of the counterdiabatic operator into nested commutators, whereas higher order terms only provide small corrections. \replace{This point is crucial because it makes the use of CRA experimentally viable as the locality of the approximate CD operator is low.}{This point is crucial because, if confirmed in general, it could make CRA experimentally viable for Ising-like models, where the locality of low-order nested commutators is low.} \textit{(iii)}~The cost of the approximate CD driving~\cite{campbell:cost-cd-1} in CRA scales as a power law of the system size, almost independently of the annealing path and on the fraction of spin-ups of the initial state. In the short-time regime, the scaling of the spectral gap as a function of $N$ in the presence of first- or second-order quantum phase transitions seems not to be a decisive point to determine the performance of CRA. \textit{(iv)}~By studying the time to solution and the ground-state fidelity as a function of the annealing time, we showed that CRA always outperforms ARA and reduces to the latter when the dynamics are adiabatic. Our results are qualitatively independent of the value of the transverse field and of the exponent $p$, see Appendices~\ref{app:gamma-2} and~\ref{app:larger-p}.

As mentioned in the main text, also in the (alternative) iterated reverse annealing protocol the initial state is classical and is supposedly close in Hamming distance to the target state. However, in IRA \textit{diabatic} processes~\cite{Crosson2021} are exploited to enhance the overlap with the target state, in a process that can also be iterated using the output state of the routine as the new input state. In this case, counterdiabatic driving would be detrimental for reverse annealing as it would trap the system in the state where it was initialized. By contrast, we showed that CD driving can be beneficial for ARA since the latter relies on the adiabatic principle to keep the system close to its ground state at all times. To further remark the differences between ARA and IRA in a different context, Ref.~\cite{passarelli:reverse-ira} showed that dissipation in the form of collective and independent dephasing is helpful for convergence in IRA while the opposite is true for ARA~\cite{passarelli:reverse-ara}. 
at the point that standard quantum annealing can outperform it~\cite{passarelli:reverse-ara}. Building on these previous results, it might be interesting to study the effect of dissipation on the CRA protocol, compared with counterdiabatically-driven quantum annealing. On the one hand, we expect that, in the weak coupling limit, dissipation will not play a major role. On the other hand, this analysis might be interesting in view of the recently proposed variational approach to CD driving open quantum systems~\cite{passarelli:cd-open} which could allow to define an open-system generalization of CRA. We leave this analysis to future works.

\begin{acknowledgements}
This work has been funded by project code PIR01\_00011 “IBiSCo”, PON 2014-2020.
\end{acknowledgements}

\appendix

\section{\texorpdfstring{$\Gamma = 2$}{Gamma = 2}}
\label{app:gamma-2}

\begin{figure}[t]
\centering
\includegraphics[width=\columnwidth]{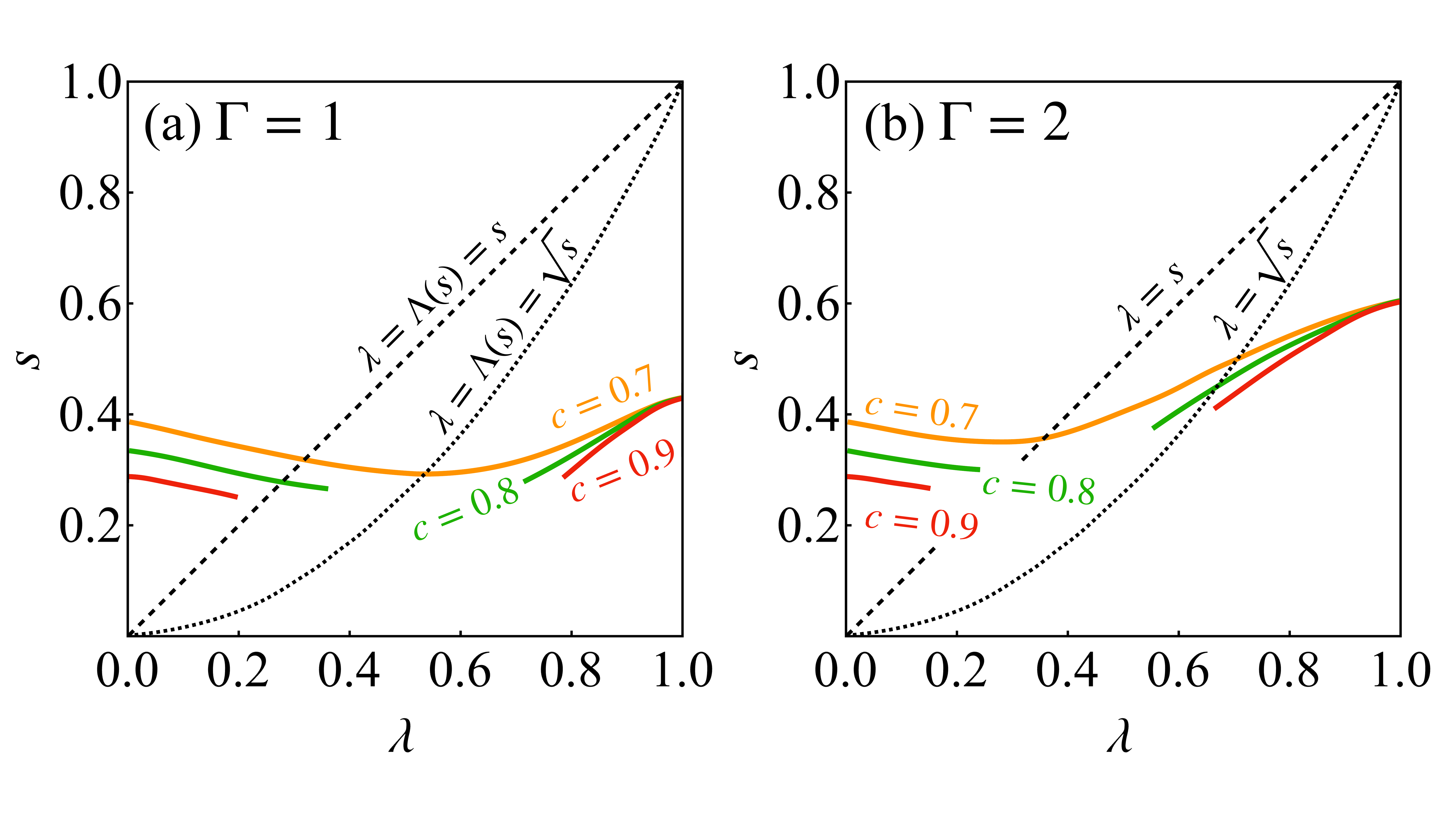}
\caption{First-order critical lines in the $(\lambda, s)$ parameter space for several values of $c$. Panel(a): $\Gamma = 1$; panel (b): $\Gamma = 2$.}
\label{fig:first-to-second-1-2}
\end{figure}

Comparing the two panels of Fig.~\ref{fig:first-to-second-1-2}, we see that by increasing the value of the transverse field $\Gamma$ we can tilt the first-order critical lines, so that the gap in the middle of the diagram moves towards smaller values of $\lambda$. Consequently, the role of the two annealing paths with $q = 1$ and $q = 1/2$ is effectively exchanged in the case of $c = 0.8$, compared to $\Gamma = 1$. Indeed, as evident from panel~\ref{fig:first-to-second-1-2}(b), for $\Gamma = 2$ the annealing path $\Lambda(s) = \sqrt{s}$ does cross the first-order critical line, whereas $\Lambda(s) = s$ does not; the opposite was true for $\Gamma = 1$. Instead, the two lines for $c = 0.7$ and $c = 0.9$ are similar to those for $\Gamma = 1$, thus for these two cases we will repeat the analysis of the previous section considering the annealing path with $q = 1$. Given our findings (see the main text), the spectral properties of the ARA Hamiltonian do not affect the performance at short times.

\begin{figure}[t]
\centering
\includegraphics[width=\columnwidth]{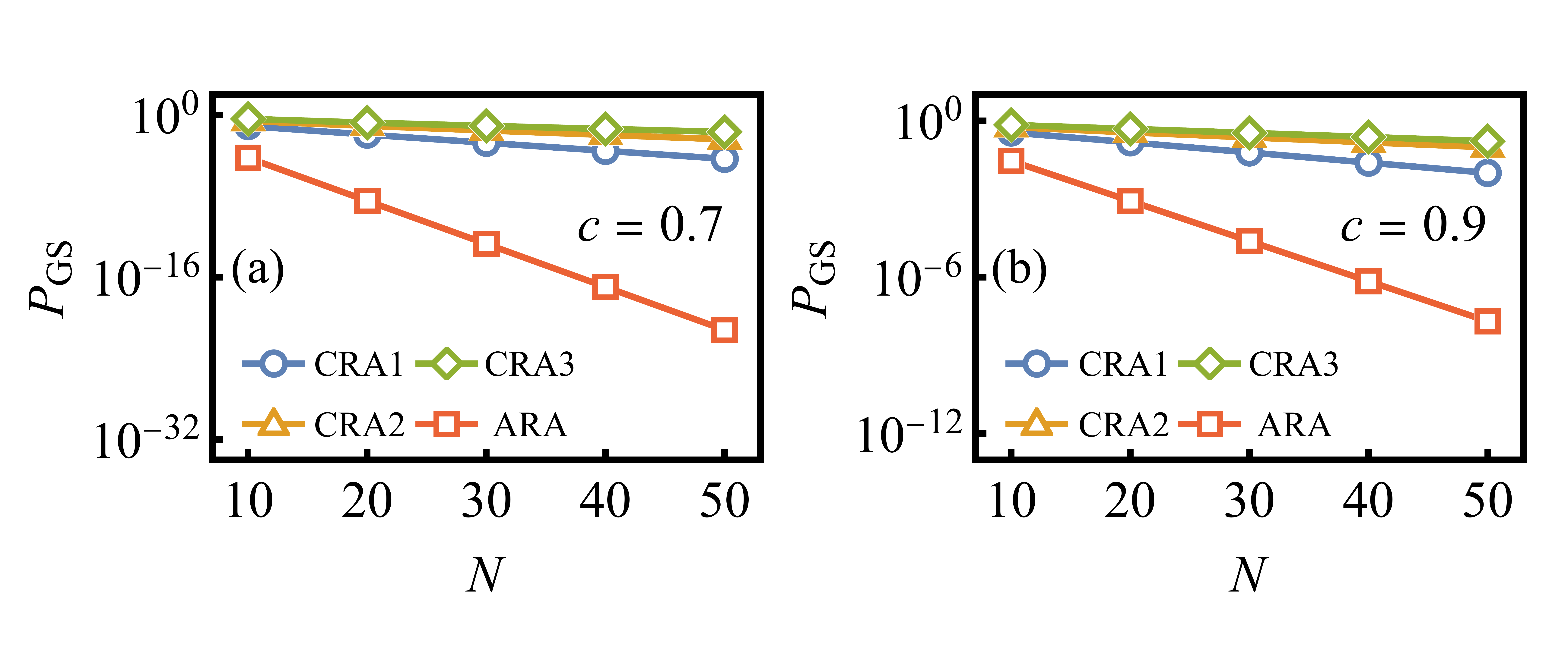}
\caption{Ground-state fidelity versus $N$, $p = 3$, for $\Lambda(s) = s$, $\Gamma = 2$, $\tau = 1$. Panel (a): $c = 0.7$. Panel (b): $c = 0.9$.}
\label{fig:c-0.7-0.9-gamma-2}
\end{figure}

In Fig.~\ref{fig:c-0.7-0.9-gamma-2}(a), we plot $P_\text{GS}$ as a function of the system size $N$ for $c = 0.7$. The same considerations of Sec.~\ref{sec:results} apply here as well: in the absence of CD driving, the GS fidelity at the end of the ARA protocol decays exponentially as a function of $N$. Moving to CRA, the scaling remains exponential but the scaling exponent becomes smaller. Also in this case, we observe an overall saturation of the performance of CD driving already at order $K=2$, which suggests that our results are robust with respect to changes in the Hamiltonian parameters. This is also confirmed by the similar results that we observe for the case of $c = 0.9$, shown in Fig.~\ref{fig:c-0.7-0.9-gamma-2}(b). The fitted scaling exponents are shown in Tab.~\ref{tab:exponents-gamma-2}. Comparing with Tab.~\ref{tab:exponents}, we notice that the stronger transverse field improves the scaling of $P_\text{GS}$ with $N$ for both ARA and CRA. These results are also in good agreement with Eq.~\eqref{eq:pgs-perturbative} in the case of ARA.


\begin{table}[tb]
    \centering
    \begin{tabular}{l@{\hskip 2ex}c@{\hskip 2ex}c}
    \toprule
          & $c = 0.7$ & $c = 0.9$\\
    \midrule
    ARA & 1.40    & 0.51 \\
    CRA1   & 0.30    & 0.12 \\
    CRA2   & 0.16    & 0.07 \\
    CRA3   & 0.11    & 0.05 \\
    \bottomrule
    \end{tabular}
    \caption{Exponent $\gamma$ in the scaling law $P_\text{GS} \sim 2^{-\gamma N}$ for $c = 0.7$, $0.9$, with and without counterdiabatic corrections ($\Gamma = 2$).}
    \label{tab:exponents-gamma-2}
\end{table}

\begin{figure}[b]
    \centering
    \includegraphics[width=\columnwidth]{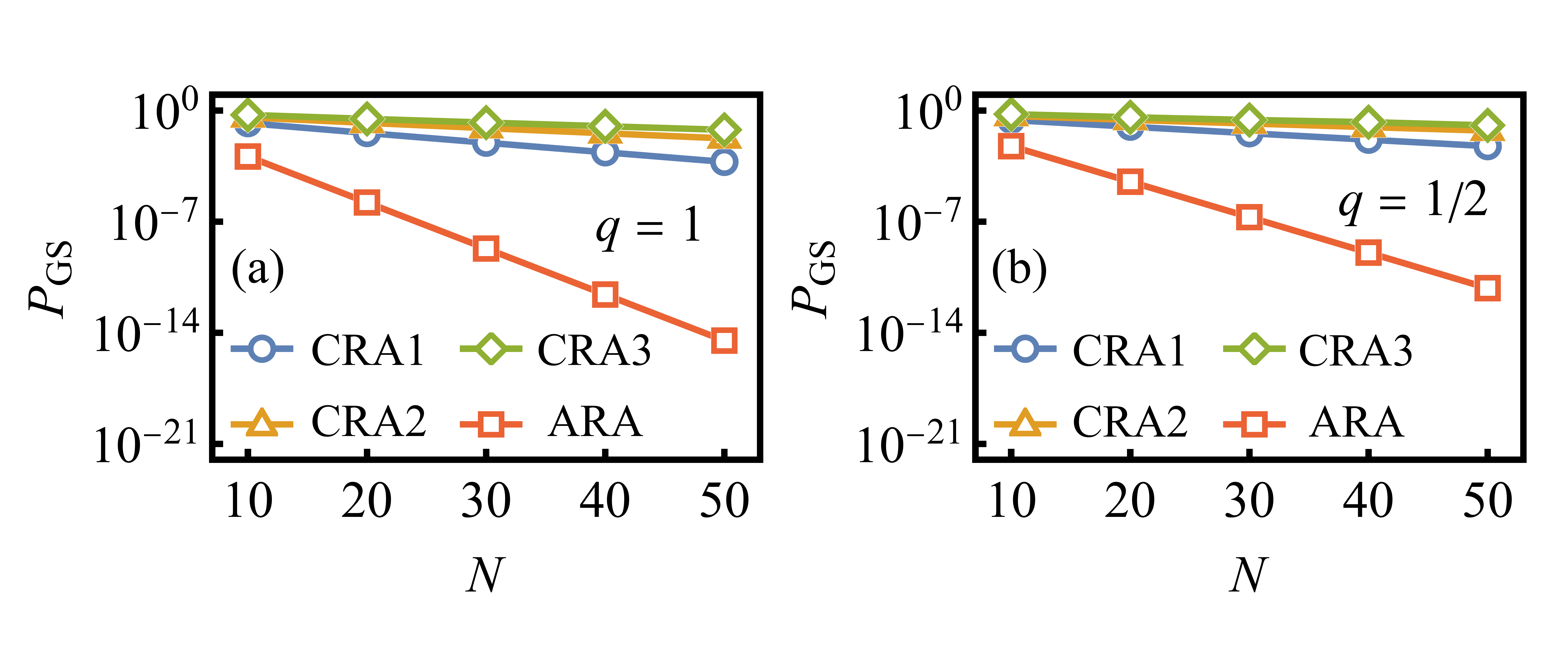}
    \caption{Ground-state fidelity versus $N$, $c = 0.8$, $p = 3$, $\Gamma=2$. Panel (a): $q = 1$; Panel (b): $q = 1/2$.}
    \label{fig:c-0.8-gamma-2}
\end{figure}


Finally, we consider the case $c = 0.8$ and compare the two annealing paths, $q = 1$ and $q = 1/2$. We show our results concerning the scaling of the ground-state fidelity with $N$ in Fig.~\ref{fig:c-0.8-gamma-2}. Here, we see something that undoubtedly confirms that the gap structure does not matter at short times: in this regime, we see indeed that the path with $q = 1/2$ is advantageous compared with $q = 1$ in terms of the exponent of the scaling relations of $P_\text{GS}$ with $N$ (see also Tab.~\ref{tab:exponents-0.8-gamma-2}), despite the fact that it crosses the first-order critical line. We find that the energetic costs of CD driving along the two annealing paths are quantitatively similar to each other (not shown).


\begin{table}[tb]
    \centering
    \begin{tabular}{l@{\hskip 2ex}c@{\hskip 2ex}c}
    \toprule
          & $q = 1$ & $q = 1/2$\\
    \midrule
    ARA & 0.67    & 0.51 \\
    CRA1   & 0.15    & 0.10 \\
    CRA2   & 0.08    & 0.06 \\
    CRA3   & 0.06    & 0.04 \\
    \bottomrule
    \end{tabular}
    \caption{Exponent $\gamma$ in the scaling law $P_\text{GS} \sim 2^{-\gamma N}$ for $c = 0.8$, for $q = 1$ and $q = 1/2$ ($\Gamma=2$).}
    \label{tab:exponents-0.8-gamma-2}
\end{table}

\begin{figure}[t]
    \centering
    \includegraphics[width=\columnwidth]{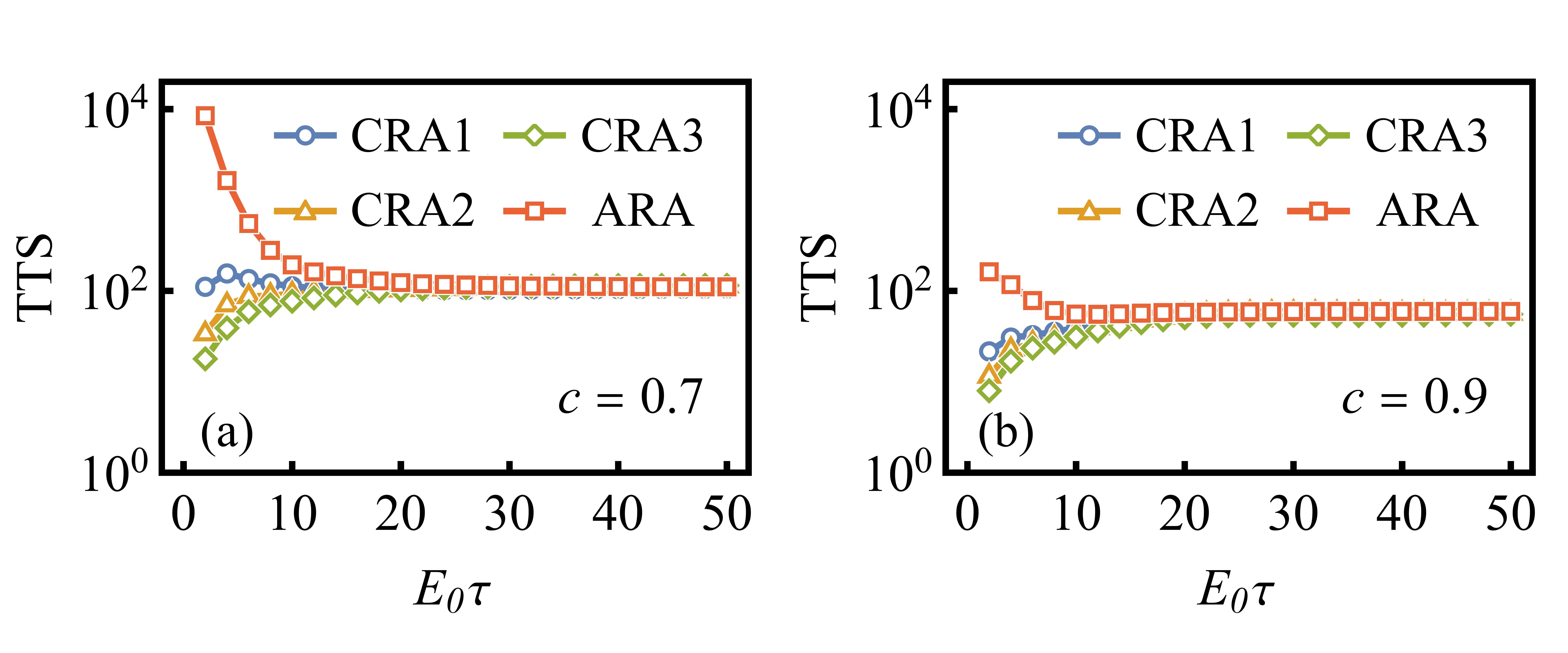}
    \caption{Time to solution\addition{, in units of $1/E_0$}, versus annealing time, for $N = 10$, $\Gamma = 2$, $q = 1$. Panel (a): $c = 0.7$; Panel (b): $c = 0.9$.}
    \label{fig:tts-10-gamma-2}
\end{figure}

In Fig.~\ref{fig:tts-10-gamma-2}, we plot the time to solution as a function of the annealing time, for $N = 10$ and $q = 1$. We consider two fractions $c$, $c = 0.7$ and $c = 0.9$. We can see the same qualitative features found also for $\Gamma = 1$, which shows that we can expect CRA to significantly outperform ARA at short times, see for instance Fig.~\ref{fig:tts-10-gamma-2}(a), where the CRA3 solution has an advantage of three orders of magnitude over ARA. The larger value of $\Gamma$ results in an overall shorter time to solution compared with the results for $\Gamma = 1$ but, despite this quantitative difference, our previous assessment remains unchanged.

\section{\texorpdfstring{$p = 5$}{p = 5}}
\label{app:larger-p}

\begin{figure}[t]
    \centering
    \includegraphics[width=0.5\columnwidth]{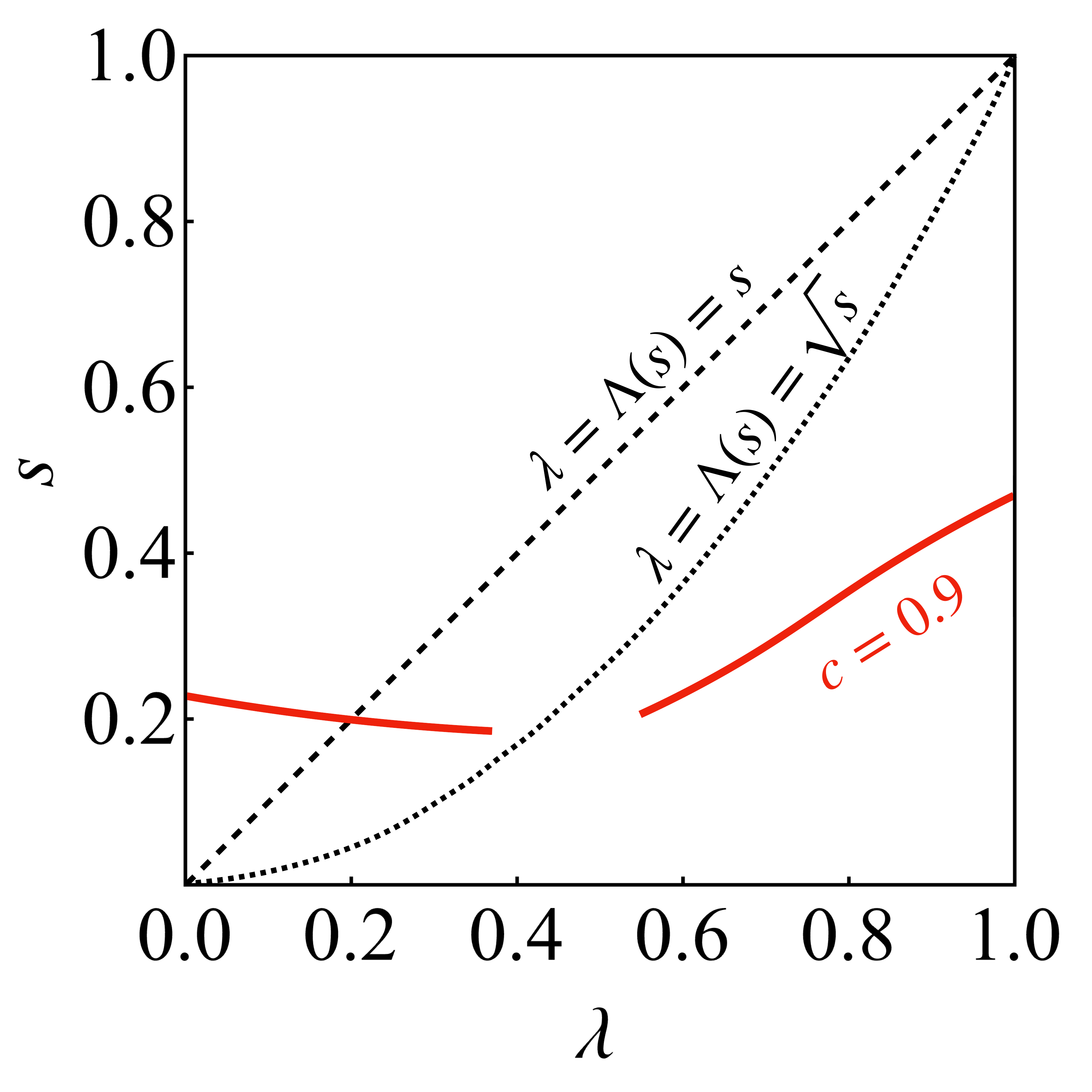}
    \caption{First-order critical lines in the $(\lambda, s)$ parameter space for $c = 0.9 > c^*$ ($p = 5$, $\Gamma = 1$).}
    \label{fig:first-to-second-ara-5}
\end{figure}

In this Appendix, we focus our attention on the case $p = 5$. For this value of $p$ the critical fraction $c^*$ above which the first-order critical line breaks is $c^* = 0.89$ ($\Gamma=1$). The linear annealing path with $q = 1$ always crosses the critical line for the values of $c$ considered in this manuscript, whereas the annealing path $\Lambda(s) = \sqrt{s}$ with $q = 1/2$ avoids the first-order transition for $c = 0.9$. This is shown in Fig.~\ref{fig:first-to-second-ara-5}.

\begin{figure}[t]
\centering
\includegraphics[width=\columnwidth]{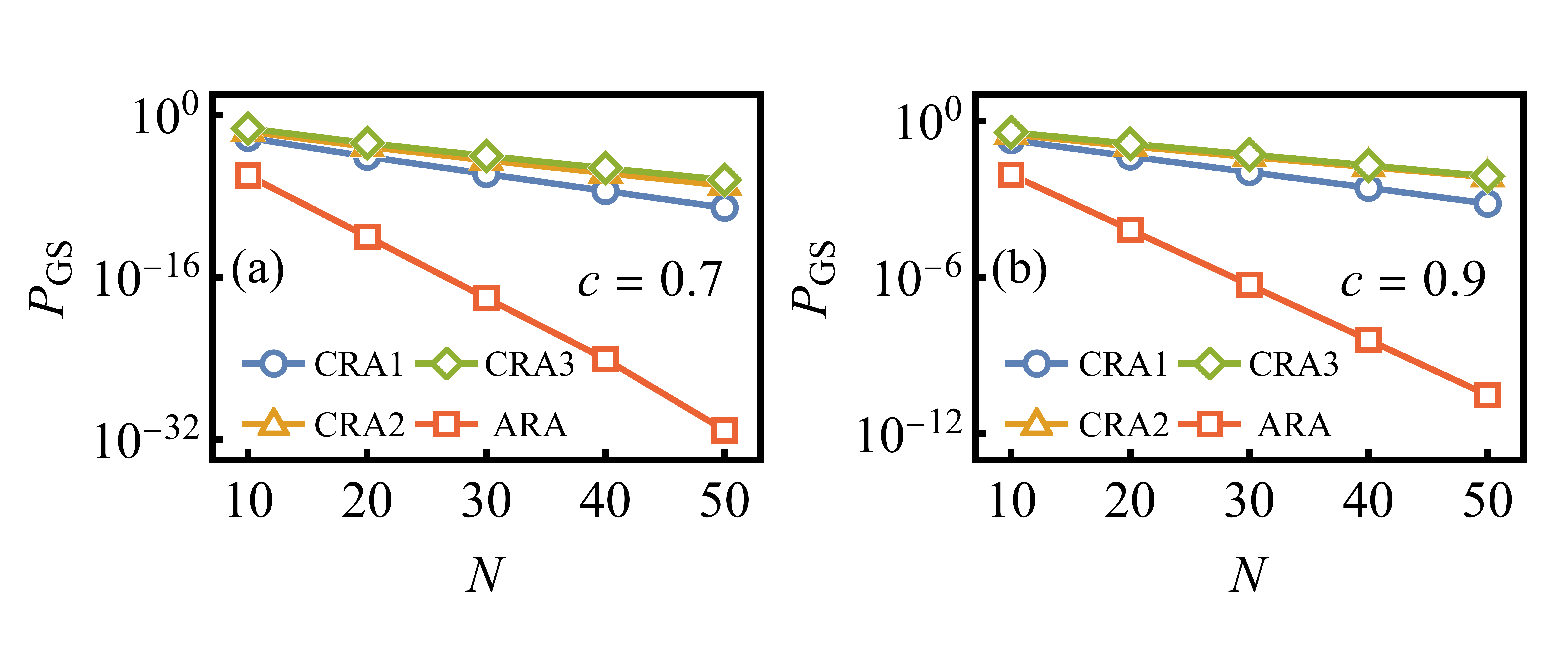}
\caption{Ground-state fidelity versus $N$, $p = 5$, for $\Lambda(s) = s$, $\Gamma = 1$, $\tau = 1$. Panel (a): $c = 0.7$; panel (b): $c = 0.9$.}
\label{fig:c-0.7-0.9-p-5}
\end{figure}

Our numerical investigation shows that the results for $p = 5$ are very similar to the ones for $p = 3$. As an example, we report in Figs.~\ref{fig:c-0.7-0.9-p-5} the scaling of the ground state probability as a function of $N$ for $c = 0.7$ [panel (a)] and $c = 0.9$ [panel (b)]. The similarity with $p=3$ is striking, at all orders and even for the uncorrected ARA protocol and is a byproduct of the perturbative analysis carried out in the main text, according to which only the number of spin flips that connect the starting state to the target state is important when the annealing time is short. Thus, we suggest that the results discussed in the main text are representative of the $p$-spin model for odd $p$.

%

\end{document}